\documentclass[journal]{vgtc}                     

\graphicspath{{figures/}{pictures/}{images/}{./}} 

\usepackage{times}                     

\usepackage{tabu}                      
\usepackage{booktabs}                  
\usepackage{lipsum}                    
\usepackage{mwe}                       
\usepackage{xspace}
\usepackage{mathptmx}  
\usepackage[normalem]{ulem} 
\usepackage{graphicx}
\usepackage{subcaption}
\usepackage{caption}
\usepackage{multirow}
\usepackage{transparent}
\usepackage{soul}
\usepackage{enumitem}
\usepackage{bm}
\usepackage{amssymb}
\usepackage{filecontents}
\usepackage{tikz}
\usepackage{amsmath}
\usepackage{pifont}
\usepackage{circledsteps}
\usepackage[table]{xcolor} 

\usepackage[switch]{lineno}

\usepackage{titlesec}
\titlespacing*{\section}{0pt}{2ex}{0.5ex}
\titlespacing*{\subsection}{0pt}{1.2ex}{0.3ex}
\titlespacing*{\subsubsection}{0pt}{1.2ex}{0.3ex}

\usepackage{tabularx}
\usepackage{booktabs} 
\usepackage{adjustbox}
\usepackage{makecell}
\usepackage{array}

\usepackage{mathptmx}                  

\onlineid{0}

\vgtccategory{Research}

\preprinttext{}

\newif\ifshowrevisions
\showrevisionsfalse    

\newcommand{\rev}[1]{%
  \ifshowrevisions
    \textcolor{cyan}{#1}%
  \else
    #1%
  \fi
}

\newcommand{\cmark}{\checkmark} 
\newcommand{\xmark}{\texttimes} 
\newcommand{\name}{SemanticXR\xspace}

\newcommand{\serverquery}{SemanticXR-SQ\xspace}
\newcommand{\localquery}{SemanticXR-LQ\xspace}

\newcommand{\baseline}{device-cloud baseline\xspace}

\pgfkeys{/csteps/fill color=black}
\pgfkeys{/csteps/inner color=white}
\pgfkeys{/csteps/inner xsep=6pt}
\pgfkeys{/csteps/inner ysep=5pt}

\title{SemanticXR: Low Power, Real-time Queryable Semantic Mapping with a Device-Cloud Architecture}

\author{%
  \authororcid{Rahul Singh}{0000-0002-3868-3906}, 
  \authororcid{Devdeep Ray}{0000-0002-3533-0064}, 
  \authororcid{Connor Smith}, and 
  \authororcid{Sarita Adve}{0000-0002-3403-5119} 
}

\authorfooter{
  \item
    Rahul Singh and Sarita Adve are with the University of Illinois
    Urbana-Champaign. E-mail: rahuls10@illinois.edu, sadve@illinois.edu.
  \item
    Devdeep Ray and Connor Smith are with NVIDIA.
    E-mail: devdeepr@nvidia.com, cosmith@nvidia.com.
}

\teaser{
\vspace{2pt}
  \centering

    \includegraphics[page=8, width=\linewidth,  height=7.8cm,keepaspectratio=false]{figures/Teaser_3options.pdf}

\vspace{-0.15in}
\caption{\name overview. \name enables real-time, open-vocabulary semantic mapping for low-power XR through a device-cloud architecture organized around objects
as first-class units of communication, execution, and memory footprint. 
Object-level innovations (green) speed up server-side mapping (\cref{sys:optimizations}) and reduce upstream bandwidth (\cref{sys:upstream}); on the device, they add a sparse local map with incremental updates and update prioritization for network-robust querying (\cref{sys:query_modes}).
Organizing every operation at object granularity allows both applications and the system to trade off resource usage against semantic quality, suiting application requirements and operating conditions respectively, without modifying the perception and mapping pipeline (\cref{sys:gen_tunable}).
}

  \label{fig:teaser}
  \vspace{-0.01in}
}

\abstract{Semantic mapping is a core service that enables grounded interactions in emerging Extended Reality (XR) applications such as AI assistants. 
Deploying this capability on mobile XR devices requires a system that is open-vocabulary, real-time, and low-power. Existing approaches are compute-intensive and assume server-class resources. Cloud offloading offers a practical path, but no existing system splits semantic mapping between the device and the cloud, and current approaches do not address how to manage communication, execution, and memory footprint across the device-cloud boundary. We present \name, the first device-cloud system for real-time, open-vocabulary semantic mapping and querying under XR power, bandwidth, and memory constraints. Our key insight is to elevate semantically identifiable objects to first-class units of system design, governing how the system communicates, executes, and manages memory across the device and the server. 
Evaluation against a new, aggressive \baseline shows that object-level system organization improves server-side mapping latency by 2.2$\times$ at equivalent semantic quality. Object-level depth-mapping co-design maintains upstream bandwidth under 2.5\,Mbps. 
On the device, an object-level sparse local map with incremental updates and update prioritization enables 
sub-100\,ms query latency for up to 10{,}000 objects even under network drops, supports tens of thousands of objects within 500\,MB memory footprint, and scales downstream bandwidth with map changes rather than total scene size. The system adds only about 2\% to idle device power.

}


\begin{document}


\firstsection{Introduction}

\maketitle




Extended Reality (XR) has the potential to transform areas such as education, healthcare, accessibility, and industrial work. To support emerging capabilities such as spatial object search, AI assistants, and context-aware scene understanding, XR devices must go beyond reconstructing 3D geometry to associating semantic meaning with the physical environment. 
This requires a semantic mapping service that incrementally builds and retains a queryable 3D map linking geometry with meaning. For example, when a user asks "Where are my keys?," the system should be able to guide them to the keys' location, even if the keys are not currently in view, and highlight the keys once they come into view.

Deploying semantic mapping as a service on XR devices imposes several concurrent requirements. From the algorithm side, because XR devices operate in diverse and previously unseen environments, the semantic layer must be open-vocabulary, supporting recognition beyond fixed categories. From the system side, the system must operate in real time and within strict power budgets to support interactive applications on battery-constrained, all-day wearable devices. The resulting system must deliver high semantic quality, but quality demands vary across applications, creating opportunities to trade off resource usage for semantic coverage and geometric detail.

Recent algorithmic work in vision and robotics has advanced open-vocabulary semantic mapping by lifting outputs from 2D foundation models into persistent 3D representations~\cite{conceptfusion, conceptgraph, clio, open-fusion, lerf, langsplat, ovir3d, onlineAnySeg, onemaptofindthemall}. Some of these approaches achieve real-time performance, but they assume server-class GPUs, with compute and power resources well beyond those available on mobile XR devices. No existing system delivers open-vocabulary, real-time semantic mapping within the power constraints of mobile XR devices.

\newlength{\nmwidth}
\setlength{\nmwidth}{\dimexpr 2.5cm+1.0cm+2.2cm+4\tabcolsep\relax} 
\begin{table*}[t]
\small
\centering
\newcolumntype{L}[1]{>{\raggedright\arraybackslash}p{#1}}
\newcolumntype{C}[1]{>{\centering\arraybackslash}p{#1}}
\begin{tabularx}{\textwidth}{@{}L{2.5cm}C{1.0cm}C{2.2cm}C{1.0cm}C{1.3cm}C{1.1cm}C{2.1cm}C{2.1cm}C{1.0cm}@{}}
\toprule
& \multicolumn{2}{c}{\textbf{Location}} & \multicolumn{6}{c}{\textbf{System Requirements}} \\
\cmidrule(lr){2-3} \cmidrule(lr){4-9}
\textbf{System Architecture} & \textbf{Mapping} & \textbf{Query} & \textbf{Device Power} & \textbf{Real-Time Mapping} & \textbf{Upstream BW} & \textbf{Query under Network Drops} & \textbf{Downstream BW} & \textbf{Device Memory} \\
\midrule
All on-device & Device & Device & \xmark & \xmark & N/A & N/A & N/A & \xmark \\
\addlinespace
\multirow{2}{2.5cm}{Straightforward device-cloud\textsuperscript{\dag}} & Cloud & Cloud & \cmark & $\sim$ & \cmark & \xmark & \cmark & N/A \\
& Cloud & Device & \cmark & $\sim$ & \cmark & \cmark & \xmark & \xmark \\
\addlinespace
\textbf{Ours: \name{} (co-designed)} & \textbf{Cloud} & \textbf{Cloud + Device} & \textbf{\cmark} & \textbf{\cmark} & \textbf{\cmark} & \textbf{\cmark} & \textbf{\cmark} & \textbf{\cmark} \\
\bottomrule
\end{tabularx}
\caption{System requirements for deploying semantic mapping under XR constraints. No single existing architecture meets all requirements for device-cloud semantic mapping under XR constraints. \name's co-designed organization, built on objects as the core abstraction for communication, execution, and memory footprint, meets all of them. $\sim$~=~partially meets; N/A~=~requirement does not apply (e.g., an all-on-device system uses no network or server). \textsuperscript{\dag}No existing system implements a device-cloud split for semantic mapping.}
\label{tab:sys_req}
\end{table*}

\begin{table*}[t]
\small
\centering
\newcolumntype{L}[1]{>{\raggedright\arraybackslash}p{#1}}
\newcolumntype{C}[1]{>{\centering\arraybackslash}p{#1}}
\begin{tabularx}{\textwidth}{@{}L{2.5cm}C{1.0cm}C{2.2cm}C{1.0cm}C{1.3cm}C{1.1cm}C{2.1cm}C{2.1cm}C{1.0cm}@{}}
\toprule
\multicolumn{3}{@{}>{\raggedright\arraybackslash}p{\nmwidth}}{\textbf{\name{} Object-Level Innovation}} & \textbf{Device Power} & \textbf{Real-Time Mapping} & \textbf{Upstream BW} & \textbf{Query under Network Drops} & \textbf{Downstream BW} & \textbf{Device Memory} \\
\midrule
\multicolumn{3}{@{}>{\raggedright\arraybackslash}p{\nmwidth}}{Object-level parallelism} & \cmark & \cmark & & & & \\
\multicolumn{3}{@{}>{\raggedright\arraybackslash}p{\nmwidth}}{Object-level geometry downsampling} & \cmark & \cmark & & & & \\
\multicolumn{3}{@{}>{\raggedright\arraybackslash}p{\nmwidth}}{Object-level depth-mapping co-design} & \cmark & & \cmark & & & \\
\multicolumn{3}{@{}>{\raggedright\arraybackslash}p{\nmwidth}}{Object-level incremental updates} & \cmark & & & \cmark & \cmark & \\
\multicolumn{3}{@{}>{\raggedright\arraybackslash}p{\nmwidth}}{Object-level sparse local map} & \cmark & & & \cmark & \cmark & \cmark \\
\multicolumn{3}{@{}>{\raggedright\arraybackslash}p{\nmwidth}}{Object-level update prioritization} & \cmark & & & \cmark & \cmark & \cmark \\
\multicolumn{3}{@{}>{\raggedright\arraybackslash}p{\nmwidth}}{Object-level configurable resource usage vs.\ quality} & \cmark & \cmark & \cmark & \cmark & \cmark & \cmark \\
\bottomrule
\end{tabularx}
\caption{\name's object-level innovations and the requirement(s) each addresses. Each is enabled by treating objects as first-class units of communication, execution, and memory footprint. Together they let \name meet every requirement in \cref{tab:sys_req}.} 
\label{tab:sys_req_innov}
\vspace{-0.2in}
\end{table*}

One on-device alternative is specialized hardware acceleration, but rapidly evolving foundation model architectures limit the effective deployment lifetime of such accelerators. Cloud offloading offers an alternate practical path: it leverages powerful server-side GPUs potentially without increasing device power, avoids dependence on custom hardware, and frees on-device resources for other latency-sensitive XR tasks~\cite{xrgo, cloudxr, remotevio, slamshare, googlestream, meshreduce, accumo, arise, elf, edgeAssistedObjectDetection, marvel, vips, rao}. However, how to partition semantic mapping across the device and the cloud under XR constraints remains an open problem.

\cref{tab:sys_req} lays out the design space and system requirements. A natural device-cloud split is to perform both mapping and querying on the server, but this leaves the device unable to answer queries during network drops, which are common in mobile XR. An alternative is to perform mapping on the server and maintain a copy of the semantic map on the device for local querying. This restores query availability during network drops, but introduces new costs: updating the device map requires transferring the full scene to the device, causing downstream bandwidth and device memory footprint to grow with scene size. On the server side, existing approaches incur high per-frame mapping latency despite access to powerful GPUs (\cref{res:real_time}). No single architecture satisfies all system requirements for deploying semantic mapping under XR constraints.

The underlying limitation is how existing approaches organize computation. Many fuse semantics into monolithic scene-level representations such as global volumetric maps~\cite{open-fusion}, tying every cost to the total scene size and making them fundamentally incompatible with device-cloud deployment. Others detect and operate on fine-grained entities such as objects~\cite{conceptgraph, conceptfusion, clio, onlineAnySeg, openmask3d}, but target algorithmic quality rather than system organization. In both cases, existing work specifies how to construct semantic maps, but not how to manage their communication, execution, and memory footprint across a device-cloud boundary.

We present \name, the first end-to-end device-cloud system that enables real-time, open-vocabulary semantic mapping and querying within the power, bandwidth, and memory constraints of mobile XR.\footnote{Although we refer to cloud offloading, \name also applies to other split-compute configurations; e.g., offloading to a puck or edge server.}  Our key insight is to elevate semantically identifiable objects to first-class units of device-cloud system design, governing communication, execution, and memory footprint across the device and the server. This object-level system organization addresses the gaps identified in \cref{tab:sys_req}, and is not tied to a specific foundation model, generalizing across pipelines that produce per-object mapping representations (\cref{disc:generalize}).

\cref{tab:sys_req_innov} summarizes \name's object-level system innovations and the system requirements they address. Object-level parallelism and geometry downsampling manage server-side computation at object granularity, improving real-time mapping latency over frame- or scene-level execution (\cref{sys:optimizations}). Object-level depth-mapping co-design downsamples depth before transmitting to the server and mitigates quality loss through per-object mapping decisions, providing a lightweight alternative to compression techniques~\cite{adaPang} for reducing upstream bandwidth with negligible device-side overhead (\cref{sys:upstream}). On the device, an object-level sparse local map bounds device memory and downstream bandwidth while enabling queries under network drops (\cref{sys:query_modes}). Object-level incremental updates keep downstream bandwidth proportional to map changes rather than total scene size (\cref{sys:query_modes}). Object-level update prioritization further reduces device memory usage and downstream bandwidth by sending and storing only relevant object updates to the device. 
Object-level configurable resource usage vs.\ quality trade-offs unify these innovations, allowing both applications and the system to adapt semantic mapping behavior to application requirements and operating conditions respectively, without modifying the perception and mapping pipeline (\cref{sys:gen_tunable}).

Together, these innovations lead to the following contributions:
\begin{enumerate}[leftmargin=*,nosep]
\item We introduce \name, the first system to enable real-time, open-vocabulary semantic mapping and querying within the power, bandwidth, and memory constraints of mobile XR devices.
\item We identify objects as the core system abstraction for device-cloud semantic mapping, elevating them to first-class units of communication, execution, and memory footprint management. This abstraction enables the innovations~(\cref{tab:sys_req_innov}) that collectively address all the gaps identified in \cref{tab:sys_req}.
\item We enable per-object configurable resource usage vs.\ quality trade-offs, allowing diverse applications and the system to adapt semantic mapping to application requirements and operating conditions respectively, without modifying the perception and mapping pipeline.
\end{enumerate}

Since no existing system implements device-cloud semantic mapping, we construct a \baseline that uses the same perception models and mapping algorithm as \name but does not organize system operations at object granularity. This controlled comparison ensures that observed differences are attributable to system design rather than algorithmic or model choice. As discussed, approaches that fuse semantics into monolithic representations remain architecturally incompatible with device-cloud deployment (\cref{sec:relatedwork}). 
Among compatible approaches, our device-cloud baseline is the only one to achieve real-time mapping latency without sacrificing semantic quality: the only other real-time approach has worse quality, and every approach with comparable quality runs offline (taking seconds to minutes per frame) (\cref{res:real_time}).
Over this \baseline, \name improves real-time mapping latency by 2.2$\times$ at equivalent semantic quality, maintains upstream bandwidth under 2.5\,Mbps, and enables sub-100\,ms query latencies even under network drops while supporting tens of thousands of objects within 500\,MB. Downstream bandwidth scales with map changes rather than total scene size, and the system adds only about 2\% to idle device power. 

The \name implementation is available as open source~\cite{nvidiaXRAI,illixrWebsite} and a demonstration video is available as supplementary material.

\section{Background}
\label{sec:background}

\subsection{Geometric Mapping and Semantic Mapping}
\label{geometric_semantic}
Geometric mapping reconstructs a 3D representation of the environment, typically as a mesh, point cloud, or volumetric model, using inputs such as device pose, depth, and RGB frames. These maps support core XR functionalities such as collision detection, occlusion handling, and spatial audio. However, geometric mapping alone cannot assign semantic meaning to the scene; for example, it cannot distinguish or track objects or determine their semantic attributes.

Many emerging XR applications require understanding not only \textit{where} surfaces exist, but also \textit{what} they represent and how they relate to one another. Supporting such applications requires semantic mapping, which augments geometric reconstructions with persistent, queryable, and spatially grounded semantic attributes. Unlike per-frame semantic perception, semantic mapping maintains a persistent map with spatio-temporal consistency, recognizing previously observed objects, associating new observations with existing entities, and updating their attributes in place rather than creating a new map entry for each observation. 
Thus, realizing semantic mapping in XR requires not only accurate semantic inference, but also a system that manages a persistent semantic map under power, bandwidth, and memory constraints.

\subsection{Foundation Models and Open-Vocabulary Semantic Mapping}
\label{bg:foundation_models}
Foundation models have substantially advanced 2D scene understanding by enabling open-vocabulary recognition and generalization to previously unseen categories. Models for grounded object detection, segmentation, captioning, and vision-language embedding extraction~\cite{gdino, ram, sam, sam2, clip, openclip1, openclip2} provide strong building blocks for semantic mapping in XR. However, these models produce view-centric 2D predictions and do not natively maintain open-vocabulary semantics over persistent 3D scenes. Recent work addresses this by lifting outputs from 2D foundation models into persistent 3D representations, as described in \cref{bg:semantic_mapping}.

\subsection{Semantic Mapping using 2D Foundation Models}
\label{bg:semantic_mapping}
Semantic mapping using 2D foundation models has been explored in several prior works~\cite{conceptfusion, conceptgraph, open-fusion, ovir3d, multimodal3Dfusion:ISMAR24, onemaptofindthemall, onlineAnySeg, 3d_mem}. These approaches differ in how they structure semantic information. Some embed 2D region semantics~\cite{regionclip,seem} into a monolithic 3D representation~\cite{open-fusion}, tying all costs to total scene size. Others organize semantics around identifiable objects~\cite{conceptfusion, conceptgraph}, producing discrete per-object representations. Both families target mapping quality rather than system organization and mobile XR constraints; neither addresses how the semantic map is managed across a device-cloud boundary.

All the above approaches rely on computationally intensive foundation models, requiring server-class GPUs well beyond the power budget of mobile XR devices. As discussed in the introduction, monolithic representations that tie every cost to total scene size are fundamentally incompatible with device-cloud deployment. We therefore build on the object-based family of pipelines, which produce discrete per-object representations. Our contributions are not tied to a specific foundation model but apply to any pipeline that produces per-object representations (\cref{disc:generalize}).

\subsubsection{Semantic Mapping Flow}
\begin{figure}[h]
    \centering
    \includegraphics[width=1\columnwidth, trim={0 16 0 10},clip]{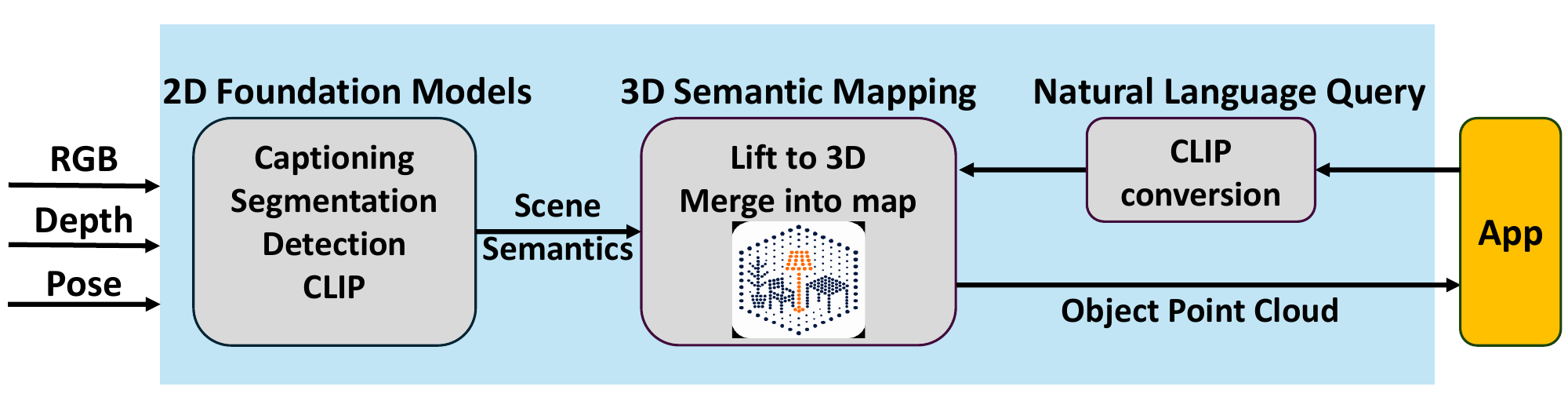}
    \caption{Representative semantic mapping pipeline using 2D foundation models. Per-frame semantic predictions are lifted into 3D using depth and pose, aggregated into a semantic map, and exposed through a query interface.}
    \label{fig:highlevel}
\vspace{-0.2in}
\end{figure}

\cref{fig:highlevel} illustrates a representative semantic mapping pipeline. At a high level, the pipeline consists of two stages. First, per-frame semantic information is extracted using open-vocabulary models, producing per-object predictions. Second, these predictions are lifted into 3D using depth and camera pose and incrementally associated with existing objects in the map based on spatial and semantic similarity. Transient observations are pruned over time to mitigate noise.

\subsubsection{Querying the Semantic Map}
\label{bg:queries}

Once a semantic map is constructed, \rev{users issue object-grounding queries: natural-language descriptions that retrieve the best-matching scene objects} (\cref{fig:highlevel}). 
The system matches a semantic embedding of the query text against per-object descriptors (e.g., using CLIP embeddings and cosine similarity) and returns the best-matching objects along with their 3D representations, which the app can then use to highlight them in the real world. \rev{We characterize this query scope and its extension to relational, affordance, and free-space queries in \cref{disc:query_scope}.} Because queries operate over the persistent semantic map, their cost depends on how that map is maintained and, in device-cloud settings, how much of it must be retained on the device.
\section{\name}
\label{sec:system}


\name builds on the object-based semantic mapping pipelines described in \cref{bg:semantic_mapping} and illustrated in \cref{fig:highlevel}. As discussed, these approaches are largely algorithmic: they target mapping quality on powerful GPUs and are oblivious to the power, bandwidth, and memory constraints of XR devices. As summarized in \cref{tab:sys_req}, no existing approach addresses how to manage communication, execution, and memory footprint across a device-cloud boundary. \name is a device-cloud system that bridges this gap, providing real-time, open-vocabulary semantic mapping and querying for low-power XR devices.

The key insight of \name is to elevate semantically identifiable objects to first-class units of device-cloud system design, governing how semantic information is communicated, executed, and stored across the device and the server. \rev{A map object is identified by a stable object ID and consists of semantic and geometric descriptors: a semantic embedding, a class label, a 3D point cloud, and a bounding box}. This abstraction generalizes across pipelines that produce per-object representations.

As illustrated in \cref{fig:teaser}, the XR device streams synchronized RGB and depth frames of the user's physical space along with the associated device pose to the server, where the semantic mapping pipeline~(\cref{bg:semantic_mapping}) detects objects, extracts semantic embeddings, and incrementally associates observations with existing objects and adds any freshly observed objects to the map. When the network is available, queries (\cref{bg:queries}) are evaluated against the full server-side map. During a network outage, the device falls back to a local semantic map maintained through object-level updates from the server.

This object-level system organization enables the innovations summarized in \cref{tab:sys_req_innov} and detailed in the following subsections: object-level parallelism and geometry downsampling for improved server-side mapping latency (\cref{sys:optimizations}), an object-level sparse local map with incremental updates and update prioritization for network-robust querying with bounded device memory and downstream bandwidth (\cref{sys:query_modes}), depth-mapping co-design for lightweight upstream bandwidth reduction (\cref{sys:upstream}), and per-object configurable resource usage vs.\ quality trade-offs that address communication, execution, and memory footprint (\cref{sys:gen_tunable}).


\subsection{Mapping Latency}
\label{sys:optimizations}

Open-vocabulary semantic mapping composes multiple foundation models followed by incremental 3D association, resulting in substantial server-side computational load. When execution is organized at frame or scene granularity, all computation is treated uniformly regardless of per-object variation in size and complexity, limiting opportunities for parallelism and missing potential efficiency gains available from this variability.

\noindent\textbf{Object-level parallelism.} \name structures server-side execution at object granularity. After object proposals are generated for a frame, subsequent processing is performed independently for each detected object (\cref{fig:teaser}). By moving away from a frame-level execution, segmentation and vision-language feature extraction are parallelized across objects within a frame, improving GPU utilization and reducing per-frame processing latency.

\noindent\textbf{Object-level geometry downsampling.} Projecting objects into 3D produces highly variable per-object point cloud sizes, depending on the size of the object, leading to disproportionate computational cost. 
Per-object geometry is needed to place each object in 3D, both for associating and merging observations by spatial proximity and for returning an object's location in response to queries.
Because object association and merging depend on spatial proximity and semantic similarity but not on high-fidelity geometric detail, \name caps the number of points per object through geometry downsampling, bounding per-object computation and improving mapping latency without degrading semantic quality. A system parameter controls this bound, enabling applications to adjust the trade-off between geometric detail and mapping latency (\cref{tab:tunable_params}).

Together, object-level parallelism and geometry downsampling improve server-side mapping latency. 


\newcolumntype{L}[1]{>{\raggedright\arraybackslash}p{#1}} 
\newcolumntype{Y}{>{\raggedright\arraybackslash}X}

\begin{table*}[!ht]
    \centering
    \begin{tabularx}{\textwidth}{L{2.7cm} L{4.3cm} Y L{2cm}}
        \toprule
        \textbf{Configurable Trade-off} &
        \textbf{Exposed System Knobs} &
        \textbf{Effect and Trade-off} &
        \textbf{Tuned values} \\
        \midrule

        Query latency vs. device power &
        \texttt{net\_latency\_switch\_threshold} &
        Switches between \serverquery and \localquery; trades device power for latency. &
        $100$ ms \\
\addlinespace
        Mapping fidelity vs server cost &
        \texttt{skip\_mapping\_set}, \texttt{max\_object\_points\_server} &
        Selects which classes the server maps and caps their per-object point budget. &
        Empty,   $2000$ pts for everything\\
\addlinespace
        Local map geometric detail vs. memory &
        \texttt{max\_object\_points\_client} &
        Caps per-object points for all/selected classes in the on-device map. &
         $200$ pts for all classes \\
\addlinespace
        Object prioritization vs. memory and bandwidth &
        \texttt{priority\_class}, \texttt{distance\_threshold} &
        Prioritizes on-device objects by class and distance; evicts the lowest priority (farthest by default) at the memory budget. &
        None, NULL \\
\addlinespace
        Local map freshness vs. downstream bandwidth &
        \texttt{local\_map\_update\_frequency} &
        Controls the frequency of incremental local map updates, trades map freshness for downstream bandwidth. &
        Every $2$ frames \\
\addlinespace
        Upstream bandwidth budget &
        \texttt{min\_mapping\_bbox\_area}, \texttt{depth\_downsampling\_ratio} &
        Defers objects below a projected-area threshold and downsamples depth before transmission. &
        $2000$ px and 5$\times $downsampling\\
        \bottomrule
    \end{tabularx}
    \vspace{-0.08in}
    \caption{Application configurable trade-offs and corresponding system-level tuning knobs in \name. Tuned values correspond to the configuration used in \name.}
    \label{tab:tunable_params}
    \vspace{-0.1in}
\end{table*}

\subsection{Query Under Network Drops: Downstream Bandwidth and Device Memory Footprint}
\label{sys:query_modes}

Interactive semantic queries in XR must remain responsive even during network drops. \name supports two query modes: \textit{Server Querying (\serverquery)}, where queries are evaluated against the full server-side semantic map, and \textit{Local Querying (\localquery)}, where queries are executed on the device using a local semantic map. Enabling \localquery requires maintaining a semantic map on the device, but a full copy of the server map is impractical: device memory would grow with scene size, and keeping the local map updated would require transferring the full map, causing downstream bandwidth to grow with scene size as well. \name addresses both constraints through three object-level innovations.


\noindent\textbf{Object-level sparse local map.} \name maintains a local semantic map on the device, organized as a collection of per-object entries. \rev{Each entry stores the same fields as a server-side object, but with the point cloud further downsampled from the server-side representation (\cref{sys:optimizations}) to fit device memory constraints}. Downsampling reduces a retrieved object's geometric detail, not which objects a query retrieves. Query accuracy therefore is unaffected and the retained geometry remains sufficient for spatial localization. Because each object's geometry is capped at a configurable point budget rather than stored as a dense scene-wide reconstruction, per-object memory is fixed and total device memory grows only with the number of retained objects, not with scene complexity. An application configurable system parameter controls the point budget per object, enabling applications and the system to adjust the memory--quality trade-off (\cref{tab:tunable_params}). The number of retained objects is further bounded by update prioritization, described below.

\noindent\textbf{Object-level incremental updates.} Updates to the local map are transmitted as object-level incremental updates. \rev{Rather than transferring the full semantic map, the server sends newly created objects together with existing objects whose geometry and semantic descriptors were revised under new viewpoints, each keyed by its stable object ID}. As a result, downstream bandwidth is proportional to the number of changed objects rather than the total scene size. Updates are issued periodically and only after an object has been consistently observed across multiple frames, filtering out transient detections before they propagate to the device. A system parameter controls update frequency, enabling applications and the system to balance map freshness against downstream bandwidth (\cref{tab:tunable_params}).

\noindent\textbf{Object-level update prioritization.} \name{} prioritizes which objects are maintained on the device using a configurable set of priority classes, semantic relevance, and proximity to the user (\cref{tab:tunable_params}); for example, an application can favor task-relevant categories, nearby objects, or distant landmarks. This allows the system and applications to limit the local map to the most relevant content and balance it against device memory and downstream bandwidth. When the local map reaches its memory budget, admitting a higher-priority update evicts the lowest-priority retained objects, keeping the map within budget. When no priority classes or distance threshold is set, the system instead evicts the objects farthest from the user. The object-level abstraction makes it straightforward to add richer replacement policies in the future.

\noindent\textbf{Query mode switching.} 
During network drops, pending updates are buffered on the server and applied upon reconnection. Network quality is monitored using latency and transmission error signals from the RGB-D stream. When network latency exceeds a configurable threshold~(\cref{tab:tunable_params}), the system switches from \serverquery to \localquery. \localquery may therefore operate on a slightly stale state, but staleness is bounded by the most recent successful update.

\subsection{Upstream Bandwidth}
\label{sys:upstream}

Offloading semantic mapping requires the XR device to transmit synchronized RGB and depth frames of the user's physical space, along with the associated device pose, to the server.
While RGB can be efficiently compressed using hardware video encoders available on mobile XR devices, depth is typically produced as high-precision frames that cannot leverage the same hardware directly~\cite{adaPang}. While prior work has explored depth compression~\cite{adaPang}, \name instead asks a co-design question: how much can depth be approximated while preserving semantic quality? This leads to a lightweight alternative -- downsampling the depth frames prior to transmission and mitigating quality loss through per-object mapping decisions.

\textbf{Object-level depth-mapping co-design.} While depth downsampling substantially lowers transmission cost, it can degrade geometric detail if applied uniformly. Because the semantic map is organized around discrete objects, mapping decisions are made per object rather than per frame. Objects that occupy sufficient image area retain reliable depth even after downsampling and are incorporated immediately, while smaller or distant objects are deferred until additional observations improve depth reliability. A system parameter controls the minimum object area required for incorporating an observation, enabling applications to adjust the trade-off between upstream bandwidth and semantic quality (\cref{tab:tunable_params}). 

\subsection{Object-Level Configurable Resource Usage vs.\ Quality}
\label{sys:gen_tunable}

XR applications impose diverse requirements on semantic maps; e.g., some require broad scene coverage while others prioritize specific object categories or regions. Because \name organizes all system operations at object granularity, applications can trade off resource usage, including compute, memory, and bandwidth, against semantic quality per object or per category, without modifying the underlying mapping pipeline. \cref{tab:tunable_params} summarizes the exposed parameters and their system-level effects. This configurability cuts across all system operations: applications can adjust geometric detail vs.\ mapping latency via geometry downsampling (\cref{sys:optimizations}), control local map update frequency vs.\ downstream bandwidth (\cref{sys:query_modes}), and reduce upstream bandwidth via depth-mapping co-design while preserving semantic quality (\cref{sys:upstream}).

\section{Evaluation Methodology}
\label{sec:eval}

\subsection{\name Implementation Details}
\label{eval:impl}
We evaluate \name using the object-based semantic mapping pipeline described in \cref{bg:semantic_mapping}. Per-object semantic observations are generated using off-the-shelf zero-shot models for captioning (RAM~\cite{ram}), object detection (Grounding DINO~\cite{gdino}), instance segmentation (MobileSAM~\cite{mobilesam}), and vision-language embedding (MobileCLIP~\cite{mobileclip}). The device transmits RGB, depth, and pose to the server (\cref{sys:upstream}), where all semantic mapping is performed. During network drops, the device executes queries against its local semantic map (\cref{sys:query_modes}). We use the tuned parameters in \cref{tab:tunable_params} except when studying a specific tunable parameter (e.g., depth downsampling in \cref{res:bw_quality}).

\subsection{Device-Cloud Baseline and Evaluation Goals}
\label{eval:baseline}
Our evaluation characterizes the effect of object-level system organization on device-cloud deployment under XR constraints. Since no existing system implements device-cloud semantic mapping, we construct a controlled and competitive \baseline to isolate this effect.

Splitting semantic mapping requires managing communication, execution, and memory footprint across the device-cloud boundary. Monolithic approaches that fuse semantics into global representations~\cite{open-fusion} tie all three costs to total scene size, incurring prohibitive overhead. Other pipelines~\cite{conceptgraph, conceptfusion, clio, onlineAnySeg, openmask3d} produce discrete per-object map representations, which are a prerequisite for the object-level system organization introduced by \name. Our \baseline follows this family (\cref{fig:highlevel}) and uses identical perception models and mapping pipeline as \name (\cref{eval:impl}) but does not organize system operations at object granularity. The server processes frames without object-level parallelism or geometry downsampling. The device receives periodic full scene updates to support local querying, rather than incremental and sparse object-level transfers. Both \name and the \baseline transmit downsampled depth; the effect of depth-mapping co-design on upstream bandwidth is studied independently in \cref{res:bw_quality}. Any observed differences are therefore attributable solely to system organization.

As shown in \cref{res:real_time}, this \baseline already matches or exceeds prior open-vocabulary mapping approaches in semantic quality while achieving lower mapping latency on the evaluated (Replica) dataset~\cite{replica}.


\subsection{Evaluation System Setup}
\label{eval:eval_sys}
Our evaluation requires fine-grained power instrumentation and controlled low-power configurations to characterize system behavior under XR constraints. Commercial XR headsets do not expose these capabilities, limiting their suitability for evaluating power-sensitive system design. Therefore, for the bulk of our results (~\cref{sec:results}), following prior XR systems work~\cite{illixr,remotevio,xrgo}, we use an NVIDIA Jetson Orin \textit{configured in low-power mode} as a proxy XR device. The platform provides embedded GPU compute, configurable power envelopes, and detailed system instrumentation. 
The configuration in \cref{tab:jetson_power_conf} emulates a modern mobile XR headset by limiting GPU TPC count, operating frequencies, and power envelope to levels representative of contemporary XR headset power budgets. 
\rev{While it does not capture a headset's thermal dissipation, memory bandwidth, or display and sensor power, this configuration provides a reasonable approximation of headset-class compute and power for our
detailed evaluations. Additionally, we also prototype \name on a Meta Quest~3~\cite{quest3} headset and an Apple iPad~Pro~\cite{ipad} to demonstrate end-to-end working consumer XR systems, as detailed in \cref{ipad}.}


\begin{table}[htbp]
    \vspace{-0.05in}
    \centering
    \small
    \begin{tabular}{| c | c | c | c |}
    \hline
        \textbf{System Parameters}       & \textbf{Low Power} & \textbf{MAXN} \\ \hline
        CPU Frequency (MHz)     & 1728 & 2202 \\ \hline
        GPU Frequency (MHz)     &  1020 & 1301 \\ \hline
        CPU count               & 8  & 12 \\ \hline
        GPU TPC count           & 3 & 8 \\ \hline
        Idle Power           & 8.69W & 10W \\ \hline
        Power Cap           & \textbf{20W} & \textbf{60W} \\ \hline
    \end{tabular}
    \caption{Jetson power configurations. Low power mode emulates a modern XR headset. MAXN represents the maximum power setting supported by Jetson Orin. Low power mode caps power at 20W while MAXN can go up to 60W. Idle power in both cases is high.}
    \label{tab:jetson_power_conf}
    \vspace{-0.15in}
\end{table}

For the server side of \name we use a dedicated workstation equipped with an AMD Ryzen Threadripper 7960X CPU and
an NVIDIA RTX 6000 Ada GPU. This configuration represents a contemporary cloud-class server and is held fixed across all experiments.

To study the impact of network conditions on system behavior, we evaluate \name under three network configurations, with the client over the university WiFi network in all cases: (1) a low-latency configuration with an average network round-trip time (RTT) measured as $\sim$20\,ms, (2) a higher-latency wide-area path at an average measured RTT of $\sim$66\,ms, and (3) complete network outage to evaluate local querying (\cref{sys:query_modes}). These settings are used consistently across experiments unless otherwise noted. 



\subsection{Dataset and Evaluation Scope}
For our Jetson-based experiments, we use the Replica dataset~\cite{replica}, as adopted by prior semantic mapping works~\cite{conceptgraph,clio}. Replica consists of diverse indoor room and office scenes and provides synchronized RGB, depth, and camera pose sequences with ground-truth semantic labels, enabling controlled and reproducible evaluation of object-level semantic queries. Our evaluation focuses on indoor scenes. Outdoor environments typically require different perception models, but since \name does not assume specific model architectures, the same system design applies when paired with suitable outdoor models. We leave empirical validation of outdoor deployment to future work.


\subsection{Evaluation Metrics}
\label{eval:metrics}

Our evaluation quantifies how object-level system organization influences device-cloud execution efficiency, and characterizes \name's robustness to network drops.

To isolate the impact of our system design, specifically whether semantic state is managed at scene or object granularity, all experiments hold the semantic mapping pipeline constant (\cref{eval:impl}).

We evaluate \name along dimensions that correspond to the system requirements identified in \cref{tab:sys_req}: device power consumption, server-side mapping latency, upstream bandwidth, query latency under network outage, downstream bandwidth, and device memory. We additionally verify that object-level organization preserves semantic quality, so that these efficiency gains do not come at the cost of mapping accuracy. These metrics collectively characterize how \name's object-level system organization addresses the requirements of device-cloud semantic mapping (\cref{sec:results}).

\subsubsection{Latency: Semantic Mapping and Queries}
\label{eval:metrics_latency}
\textbf{Semantic mapping latency} measures the time required by the server to process a single RGB, depth, and pose frame and update the semantic map. This includes semantic perception using foundation models as well as subsequent object-level map fusion. To report throughput (Frames Per Second), we follow standard practice from the literature~\cite{onlineAnySeg, rtgslam, gpsslam}--- semantic maps are incrementally refined and do not require updates for every input frame, so we sample keyframes at a fixed interval and report throughput as total input frames divided by total keyframe processing time. We use a keyframe interval of 5; prior work commonly uses intervals of 10 or higher~\cite{onlineAnySeg,gpsslam}.

\noindent
\textbf{Local query latency} comprises text embedding extraction and similarity computation against one stored embedding per object~(\cref{bg:queries}); it scales with the number of objects and is independent of the per-object geometry stored on the device. Server-side query latency additionally includes the network round-trip to transmit the query and return the retrieved object's geometry.

\subsubsection{Semantic Quality}
\label{eval:metrics_accuracy}
We evaluate semantic quality as object-grounding accuracy: how well a natural-language query retrieves the correct objects from the semantic map. We follow the evaluation methodology of~\cite{conceptfusion} on the Replica dataset~\cite{replica}, which provides ground-truth semantic labels and object annotations. Ground-truth labels are used to generate text queries that are issued against the semantic map constructed by \name. Retrieved object point clouds are compared against ground-truth using mean class recall (mAcc) and frequency-weighted mean Intersection-over-Union (F-mIoU). These metrics are standard in prior work~\cite{conceptfusion,conceptgraph,clio}.

\subsubsection{Local Map Scalability}
To evaluate how the local semantic map scales with scene complexity, we measure device memory and query latency as a function of the number of objects stored on the device. Synthetic semantic maps are constructed by incrementally inserting object point clouds with associated vision-language embeddings. We evaluate map sizes starting from 80 objects (comparable to room-scale Replica scenes) to 1{,}000, 5{,}000, and 10{,}000 objects, representing increasingly large indoor environments. We additionally report results for 25{,}000 and 50{,}000 objects to characterize extreme cases. 

\subsubsection{Network Bandwidth Usage}
We measure upstream bandwidth as the average data rate required to stream RGB, depth, and pose from the XR device to the server, accounting for RGB compression and configurable depth downsampling, where the depth downsampling ratio denotes the per-dimension reduction in depth resolution (e.g., $5\times$ per dimension, or $25\times$ overall). Downstream bandwidth measures the volume of semantic map data transferred during map synchronization. 

\subsubsection{XR Device Power Consumption}

We measure power consumption on the XR client device using \textit{tegrastats}~\cite{Tegrastats}, sampled at 1\,ms intervals. Tegrastats reports power across major subsystems, including CPU, GPU, SoC, DRAM, and system I/O. All reported power measurements include the device's idle power, which is approximately 8.69\,W on the Jetson Orin in low-power mode. Except where noted, all \name experiments run in low-power mode (\cref{tab:jetson_power_conf}). We refer to querying at one query every three seconds as \textit{intermittent}, a heavy but plausible interactive rate, and to sustained querying at the maximum achievable rate as \textit{continuous}, an unrealistic worst case. We evaluate device power under the following configurations:
\begin{enumerate}[topsep=1pt,noitemsep,leftmargin=1em]
    \item \textbf{On-device semantic mapping and querying}: Measures total power when the semantic mapping pipeline executes entirely on the device in maximum power mode, an unrealistic deployment that we report only to quantify the power cost that \name's cloud offloading eliminates.

    \item \textbf{Server-side mapping with intermittent querying}: Measures device power during normal operation, when semantic mapping runs on the server and the device streams sensor data and issues intermittent queries (\serverquery).

    \item \textbf{Intermittent local querying under network outage}: Measures device power when the device answers intermittent queries (at a heavy rate) against its local map during a network drop (\localquery).

    \item \textbf{Intermittent local querying at scale}: Measures how device power under intermittent \localquery grows with the number of objects in the local map, using the synthetic local maps from the local-map scalability experiment.
    
    \item \textbf{Continuous (unrealistic) local querying}: Although users do not issue queries continuously, some applications may generate short bursts of quick queries. We report continuous querying as a worst-case upper bound on device peak power during a network outage, not a typical usage pattern.
\end{enumerate}

\section{Results}
\label{sec:results}
This section evaluates the object-level system innovations introduced by \name (\cref{tab:sys_req_innov}) along the evaluation dimensions defined in \cref{eval:metrics}. All experiments compare \name against the \baseline described in \cref{eval:baseline}; observed differences are attributable to system organization rather than algorithmic or model choice. We first establish that the \baseline is competitive with prior work in mapping latency and semantic quality. We then evaluate how object-level system organization affects server-side mapping latency (\cref{res:real_time}), query latency under varying network conditions (\cref{res:local_queries}), device memory and query latency for large local maps (\cref{res:large_maps}), downstream bandwidth (\cref{res:local_updates}), upstream bandwidth and its interaction with semantic quality (\cref{res:bw_quality}), and device power consumption (\cref{res:power}).






\subsection{Semantic Mapping Latency and Quality}
\label{res:real_time}
\cref{tab:quality_comparison} compares semantic quality and per-frame mapping latency across representative state-of-the-art open-vocabulary mapping approaches, our \baseline, and \name. Most of these approaches are offline, requiring seconds to minutes per frame. 
Our \baseline attains semantic quality competitive with the strongest of them, OpenMask3D, and exceeds all others, while being the only system that operates in real time. 
\name matches this quality while improving on the baseline in latency and on Clio-online, the only other real-time method, in the latency and quality trade-off.



\begin{table}[ht]
\centering
\resizebox{\columnwidth}{!}{%
\begin{tabular}{lccc}
\toprule
\textbf{Method} & \textbf{mAcc} & \textbf{F-mIoU} & \textbf{Latency (s/frame) $\downarrow$} \\
\midrule
ConceptFusion~\cite{conceptfusion} & 24.16 & 31.31 & Offline \\
ConceptGraphs~\cite{conceptgraph} & 38.72 & 35.82 & Offline \\
OpenMask3D~\cite{openmask3d} & 39.54 & 49.26 & Offline \\
Clio (batch)\textsuperscript{\dag}~\cite{clio} & 37.95 & 36.98 & Offline \\
Clio (online)\textsuperscript{\ddag}~\cite{clio} & N/R (bad)\textsuperscript{\ddag} & N/R (bad)\textsuperscript{\ddag} & $\sim$0.30\textsuperscript{*} \\
\midrule
Our \baseline & 37.0 & 47.23 & 0.57 \\
\textbf{\name} & 37.6 & 48.29 & 0.26 \\
\bottomrule
\end{tabular}
}
\caption{Open-set 3D semantic segmentation accuracy and per-frame mapping latency on eight Replica~\cite{replica} scenes. Our \baseline is defined in \cref{eval:baseline}. \textsuperscript{\dag}Clio-batch reports quality after offline post-processing, which the authors note is required for adequate quality on Replica. \textsuperscript{\ddag}N/R--Not reported. Clio-online does not report quality metrics; the authors note that quality degrades significantly even compared to Clio-batch without post-processing. \textsuperscript{*}Reported on different hardware; we do not explore further since quality is not acceptable. 
}
\label{tab:quality_comparison}
\vspace{-0.15in}
\end{table}

\begin{figure}[htbp]
    \centering
    \includegraphics[page=2,width=1\linewidth]{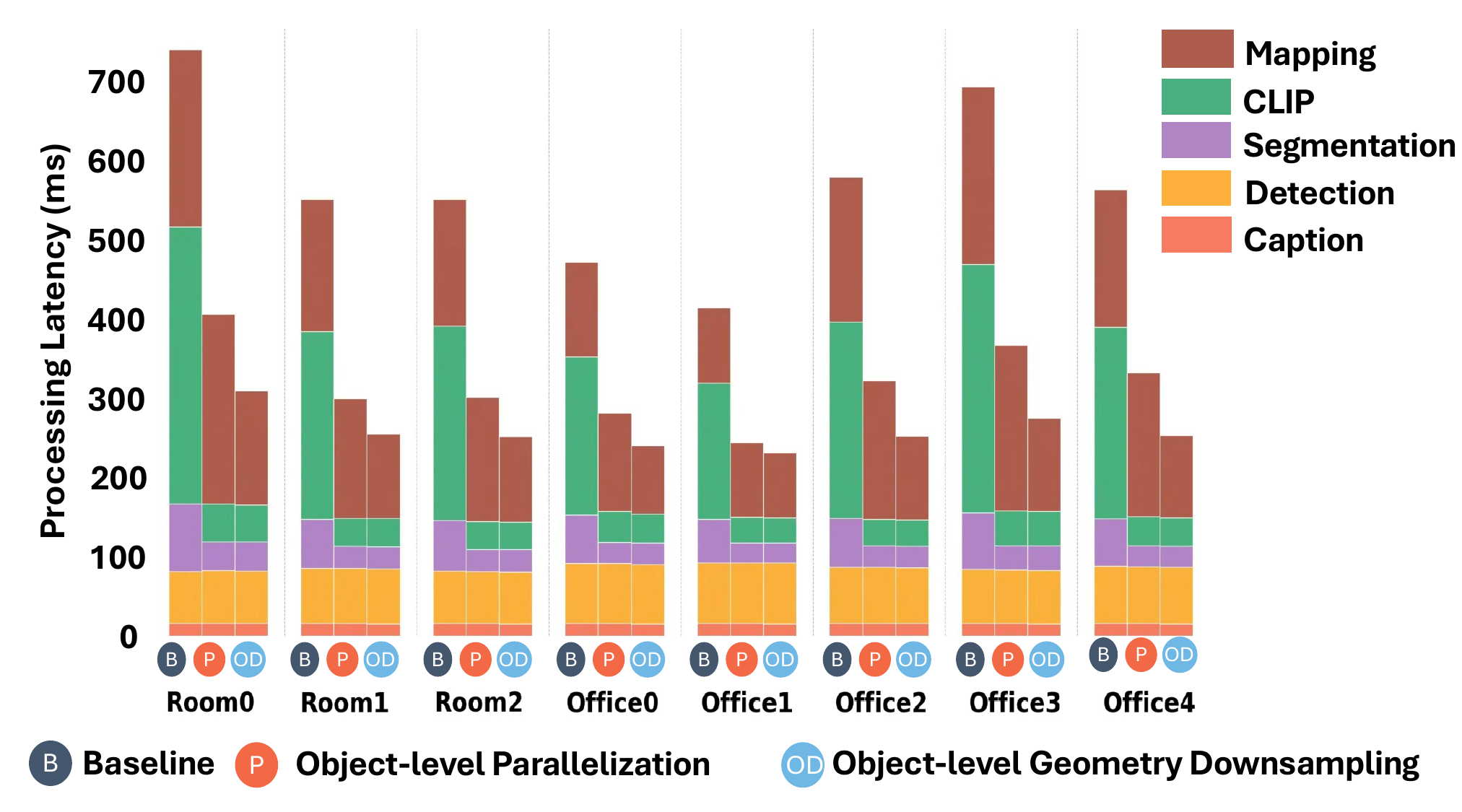}
    \vspace{-0.3in}
    \caption{Server-side semantic mapping latency across 8 scenes from the Replica dataset. 
    For each scene, the three bars form an ablation that adds one optimization at a time—\baseline (B), B + object-level parallelism (P), and B + P + object-level geometry downsampling (OD)—so each optimization's incremental latency reduction is visible. Each bar is decomposed by pipeline stage.
    }
    \label{fig:optimizations}
\vspace{-0.2in}
\end{figure}

\cref{fig:optimizations} shows how object-level parallelism and geometry downsampling reduce per-frame mapping latency over this already competitive \baseline. Object-level parallelism enables concurrent execution of segmentation and vision-language feature extraction across objects, while geometry downsampling bounds per-object point cloud size. Together, average per-frame mapping latency drops from approximately 570\,ms to approximately 260\,ms, a 2.2$\times$ improvement. Under the throughput methodology described in \cref{eval:metrics_latency}, \name achieves approximately 20\,FPS compared to approximately 8.75\,FPS for the \baseline, at equivalent semantic quality (\cref{tab:quality_comparison}).

 \begin{figure}[htb]
    \includegraphics[width=0.98\columnwidth]{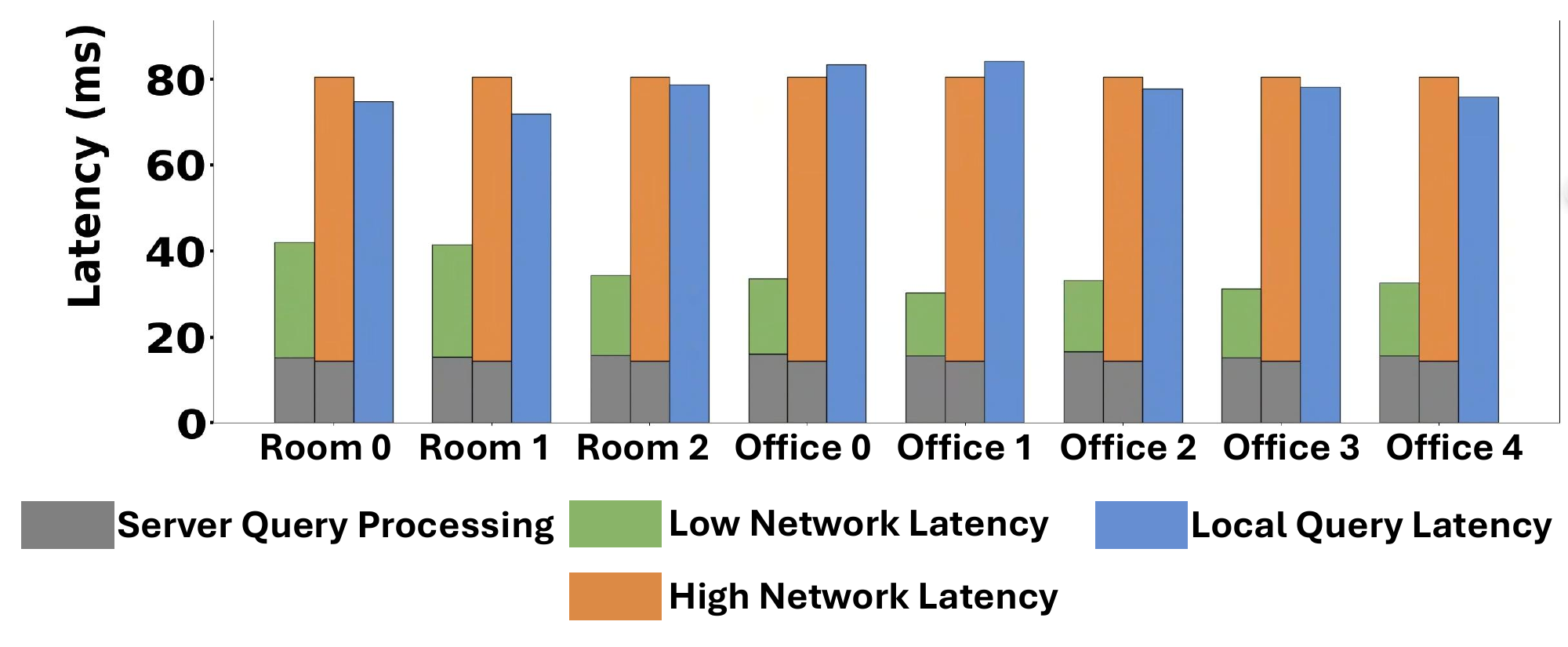}
    \caption{Average query latency for server-side queries (\serverquery), decomposed into server processing and network round-trip under the two connected network conditions, and for local queries (\localquery) under outage, across Replica scenes. Minor variations across scenes arise from differences in map size and object density.}
    \label{fig:local_query}
\vspace{-0.21in}
\end{figure}

\subsection{Network Robust Querying}
\label{res:local_queries}


\cref{fig:local_query} reports query latency across Replica scenes for server-side (\serverquery) and local (\localquery) querying under the three network conditions of \cref{eval:eval_sys}. When connected, the device uses \serverquery, which, under low network latency, is the fastest option and draws the least device power, since text embedding runs on a server-class GPU. As network latency rises, round-trip time dominates and grows variable, and \serverquery climbs to meet or exceed \localquery.

\localquery runs entirely on the device, so its latency is flat across all conditions, and it stays available during a network outage, when \serverquery cannot run. 
Because query cost depends only on the number of objects and their embeddings, and not on the geometry stored per object (\cref{eval:metrics_latency}), \localquery incurs the same query latency and device power as the \baseline's full on-device map, while using a far smaller memory footprint (\cref{res:large_maps}).


\subsection{Local Map Scalability: Device Memory and Query Latency}
\label{res:large_maps}

\begin{figure}[htbp]
    \vspace{-0.05in}
    \includegraphics[page=3,width=0.99\columnwidth]{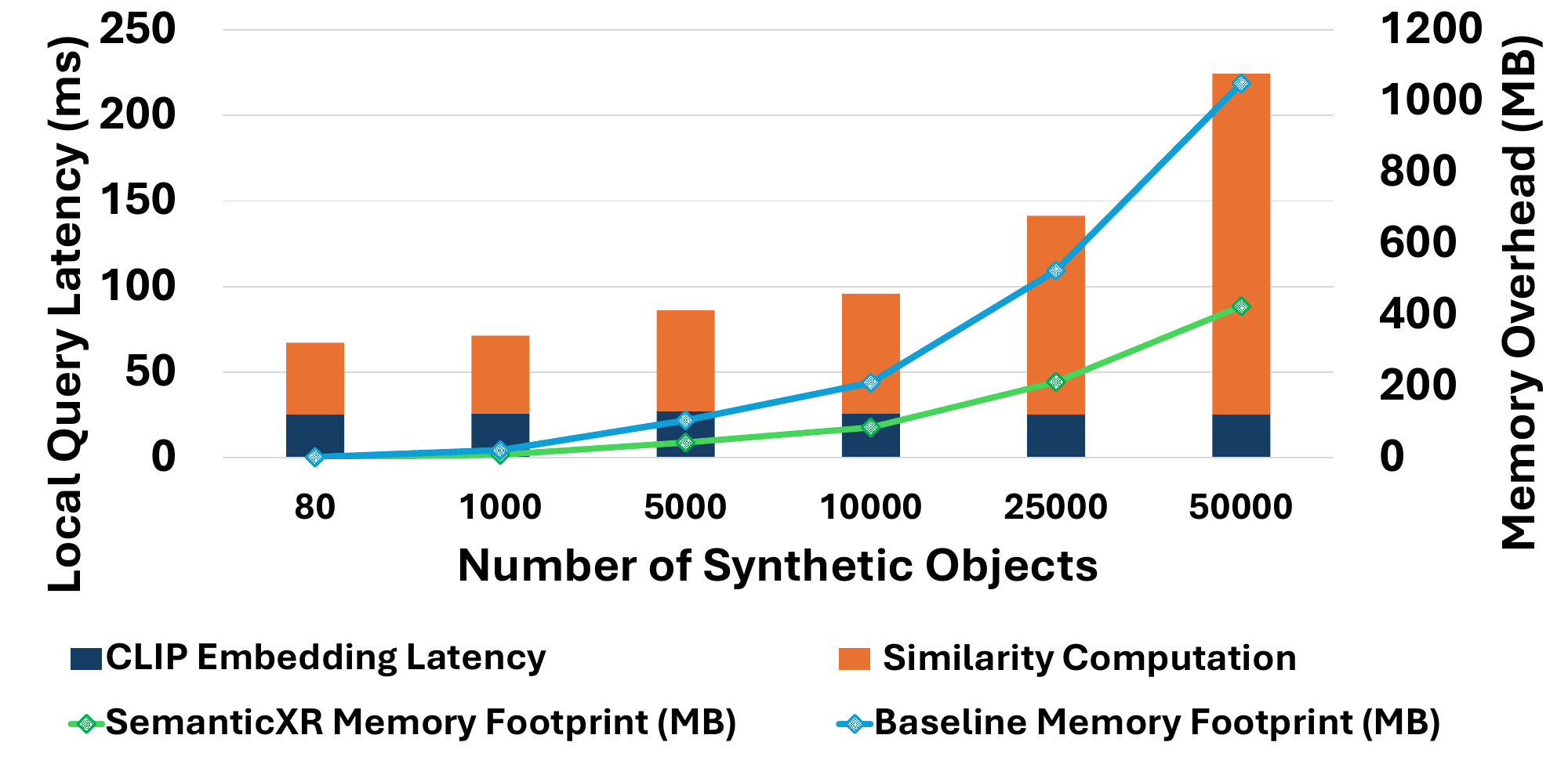}
    \vspace{-0.12in}

    \caption{Local query latency and device memory footprint as a function of the number of synthetic objects. Query latency is object-count-bound and identical for \name and the \baseline, so a single set of bars is shown. Device memory, however, grows far faster for the \baseline, which stores the full scene at full geometric resolution, than for \name's sparse per-object map.}
   
    \label{fig:synthetic_objects}
\vspace{-0.15in}
\end{figure}

We evaluate device memory footprint and local query latency as the number of objects in the local semantic map increases. \cref{fig:synthetic_objects} shows both metrics for progressively larger synthetic maps ranging from 80 objects (comparable to room-scale Replica scenes) to 50{,}000 objects. Because \name's object-level sparse local map caps per-object geometry at a configurable point budget rather than storing a dense scene-wide reconstruction (\cref{sys:query_modes}), per-object memory is bounded and total device memory grows with the number of retained objects rather than scene complexity. 

\name supports local maps containing up to 50{,}000 objects within a 500\,MB memory footprint; the associated resource-quality trade-offs are configurable via the parameters in \cref{tab:tunable_params}. \cref{fig:synthetic_objects} also shows the \baseline's device memory, which retains a dense full-resolution copy of the scene and grows far faster, exceeding 1.1\,GB at 50{,}000 objects. 
Even under the default configuration (\cref{tab:tunable_params}), which prioritizes objects by proximity with no class overrides, \name's footprint stays within the configured memory budget, so no eviction is triggered; for a scene that reaches the budget, the farthest objects are evicted to keep the map within it (\cref{sys:query_modes}).


Local query latency consists of two components: text embedding extraction, which is independent of map size, and similarity computation against stored per-object embeddings, which grows with the number of objects. For maps containing up to 10{,}000 objects, end-to-end local query latency remains below 100\,ms, enabling network-independent querying at interactive latency. Beyond 10{,}000 objects—a scale reached only in unusually large environments—query latency grows but stays acceptable.

\subsection{Downstream Bandwidth}
\label{res:local_updates}

The \baseline transmits the full semantic map to the device on every update, causing downstream bandwidth to grow with the total number of objects in the scene regardless of how many have changed. In contrast, \name transmits object-level incremental updates (\cref{sys:query_modes}), sending only newly created or modified objects.

\begin{figure}[htbp]
    \vspace{-0.14in}
    \includegraphics[page=2,width=0.99\columnwidth]{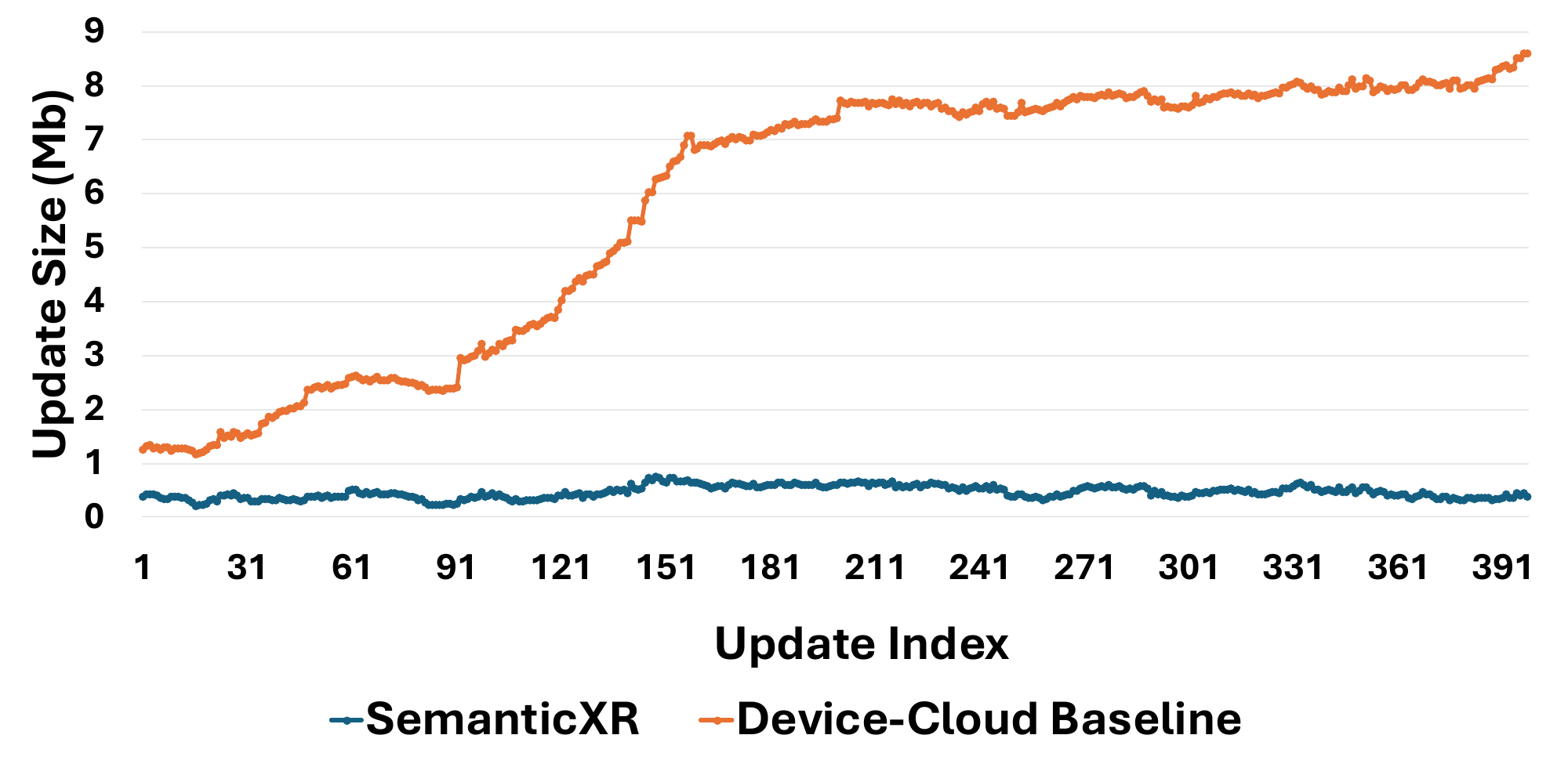}
    \vspace{-0.05in}
    \caption{Per-update downstream transfer size as a function of update index for a Replica scene. The \baseline transfers the full scene at full geometric resolution on each update, so transfer size grows with the total mapped objects until the scene is fully explored (plateau). \name's object-level incremental updates transfer only changed objects, so per-update cost remains small and decreases as exploration converges. The gap thus reflects both effects: fewer objects per update and a smaller per-object payload.}
    \vspace{-0.in}
    \label{fig:downstream_bw}
\end{figure}

\cref{fig:downstream_bw} shows this effect. The \baseline transfers the full scene at full geometric resolution on every update, so its downstream bandwidth grows with the total number of mapped objects. \name instead transfers only the changed objects, each in the sparse, downsampled form stored in its local map (\cref{sys:query_modes}), so the gap reflects both effects: fewer objects per update and a smaller per-object payload. As exploration converges and fewer new objects are discovered, \name's per-update cost decreases further, whereas the \baseline's remains at its plateau.

\subsection{Upstream Bandwidth}
\label{res:bw_quality}

As discussed in \cref{sys:upstream}, \name explores a co-design question: how aggressively can depth be downsampled before transmission while preserving semantic quality? Rather than applying compression~\cite{adaPang}, \name downsamples the depth frame and mitigates the resulting geometric loss through per-object mapping decisions that defer objects whose small projected area makes their depth unreliable until additional observations are available. \cref{tab:bw_quality} reports the resulting upstream bandwidth and semantic quality when varying depth downsampling ratio and minimum object area (\cref{tab:tunable_params}). Reducing depth resolution by $5\times$ in each spatial dimension ($25\times$ overall) decreases upstream bandwidth by approximately 90\% while causing only a small drop in quality (F-mIoU). Further reductions yield diminishing bandwidth savings while increasingly deferring object integration, as fewer objects meet the minimum area threshold per frame.

\begin{table}[h]
    \centering
    \small
    \resizebox{\columnwidth}{!}{%
    
    \begin{tabular}{| c | c | c |}
    \hline
        Depth Down-sampling  & Upstream BW (Mbps) & Quality (F-mIoU)   \\ \hline
        No Down-sampling     & 26.4 & 48.29 \\ \hline
        \(2\times\) row, \(2\times\) column (\(4\times\)) & 7.72 & 45.5 \\ \hline
        \(3\times\) row, \(3\times\) column (\(9\times\)) & 4.26 & 44.26 \\ \hline
        \(4\times\) row, \(4\times\) column (\(16\times\)) & 3.06 & 45.73 \\ \hline
         \textbf{\(5\times\) row, \(5\times\) column (\(25\times\))}  & \textbf{2.5} & \textbf{ 45.81} \\ \hline
        
    \end{tabular}
    }

    \vspace{-0.05in}
    \caption{Upstream bandwidth and semantic query quality (F-mIoU) under different depth resolutions. A $5\times$ reduction in each spatial dimension reduces bandwidth by approximately 90\% with a small impact on semantic query accuracy and is used as the default configuration. \cref{tab:quality_comparison} reports the full-resolution quality (top row).}

    \label{tab:bw_quality}
    \vspace{-0.10in}     
\end{table}






\subsection{XR Device Power Consumption}
\label{res:power}



\begin{figure}[htbp]
    \vspace{-0.10in}
    \includegraphics[page=3,width=\columnwidth]{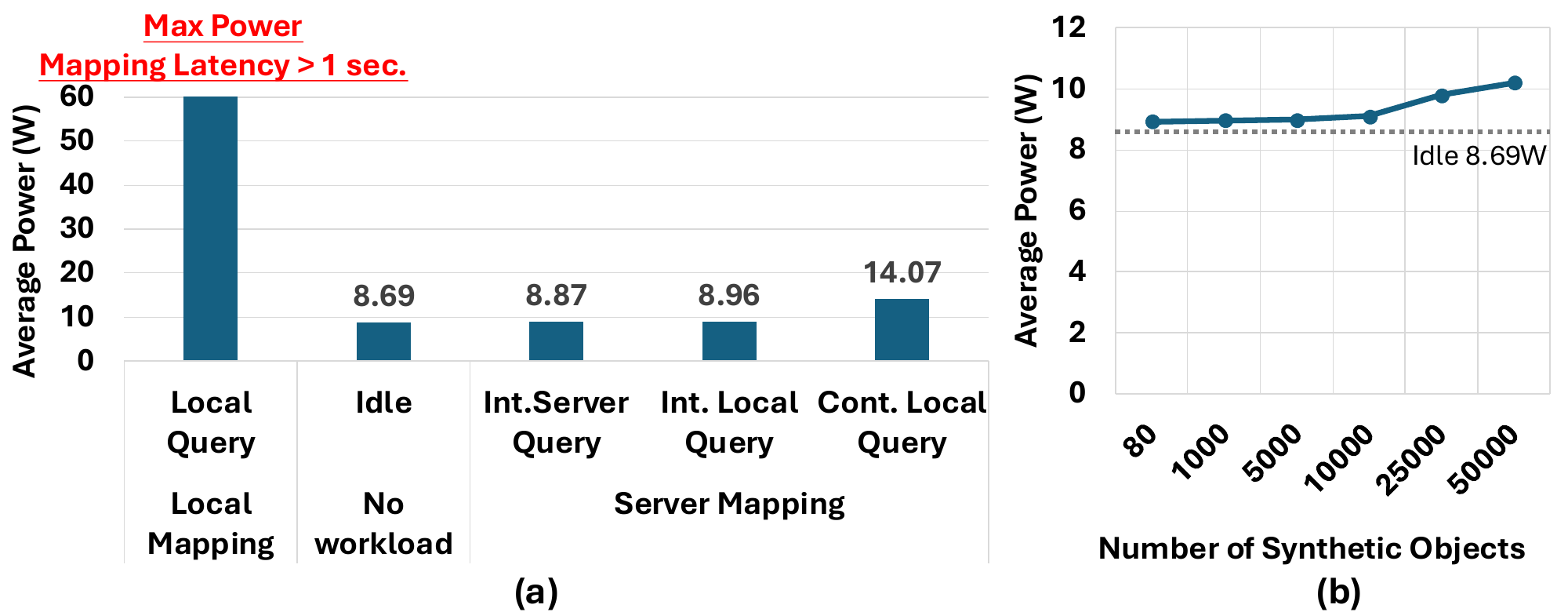}
    \vspace{-0.2in}
    \caption{XR device power on Jetson. (a) Average power by workload, grouped by mapping location and query type (Int.: intermittent; Cont.: continuous; Server/Local Query denote \serverquery/\localquery). (b) Intermittent local-query power for \name as a function of the number of synthetic objects, under network outage.}

    \label{fig:local_query_power}
\vspace{-0.15in}
\end{figure}

\cref{fig:local_query_power}(a) reports device power across \name's configurations alongside the on-device alternative it replaces. Running the full pipeline on-device draws 60\,W in maximum power mode and takes several seconds to map a single frame, confirming that on-device mapping is impractical within XR power budgets and motivating \name's cloud offloading. With mapping offloaded, normal operation (server-side mapping with intermittent \serverquery) draws 8.87\,W, only $\sim$2\% over the 8.69\,W idle power (\cref{tab:jetson_power_conf}), as the device only streams sensor data and receives results.

Under a network outage the device falls back to intermittent \localquery on its local map, drawing 8.96\,W, about 0.3\,W ($\sim$3\%) over idle. Device power here is set by CLIP embedding and similarity computation, so it grows with the number of objects rather than how the map is stored: \cref{fig:local_query_power}(b) shows it rising from 8.96\,W at 80 objects to 10.2\,W at 50{,}000, staying within XR budgets even for very large maps. To characterize worst-case behavior, we additionally measure power under continuous query execution at the maximum achievable rate of 14.7 queries per second, an unrealistic rate for user-facing XR that we report only as an upper bound. Peak power reaches 14.07\,W, remaining within the power envelope of contemporary XR devices.



\subsection{Summary}
\label{res:summary}

Object-level system organization enables real-time open-vocabulary semantic mapping with low-power XR devices. Server-side mapping runs 2.2$\times$ faster than a new aggressive baseline at equivalent semantic quality, and queries return in under 100\,ms for up to 10{,}000 objects even when the network drops. \name makes network-robust cloud offloading feasible by reducing the upstream bandwidth by 90\%, decoupling the downstream bandwidth from the scene size, and holding tens of thousands of objects under 500\,MB of memory. 
All of this costs $\sim$2-3\% power on top of idle, whether the device queries the server or its local map.


\begin{figure}[h]
    \centering
    \vspace{-0.1in}
    \begin{subfigure}[b]{0.8\columnwidth} 
        \centering
        \includegraphics[width=1.1\columnwidth]{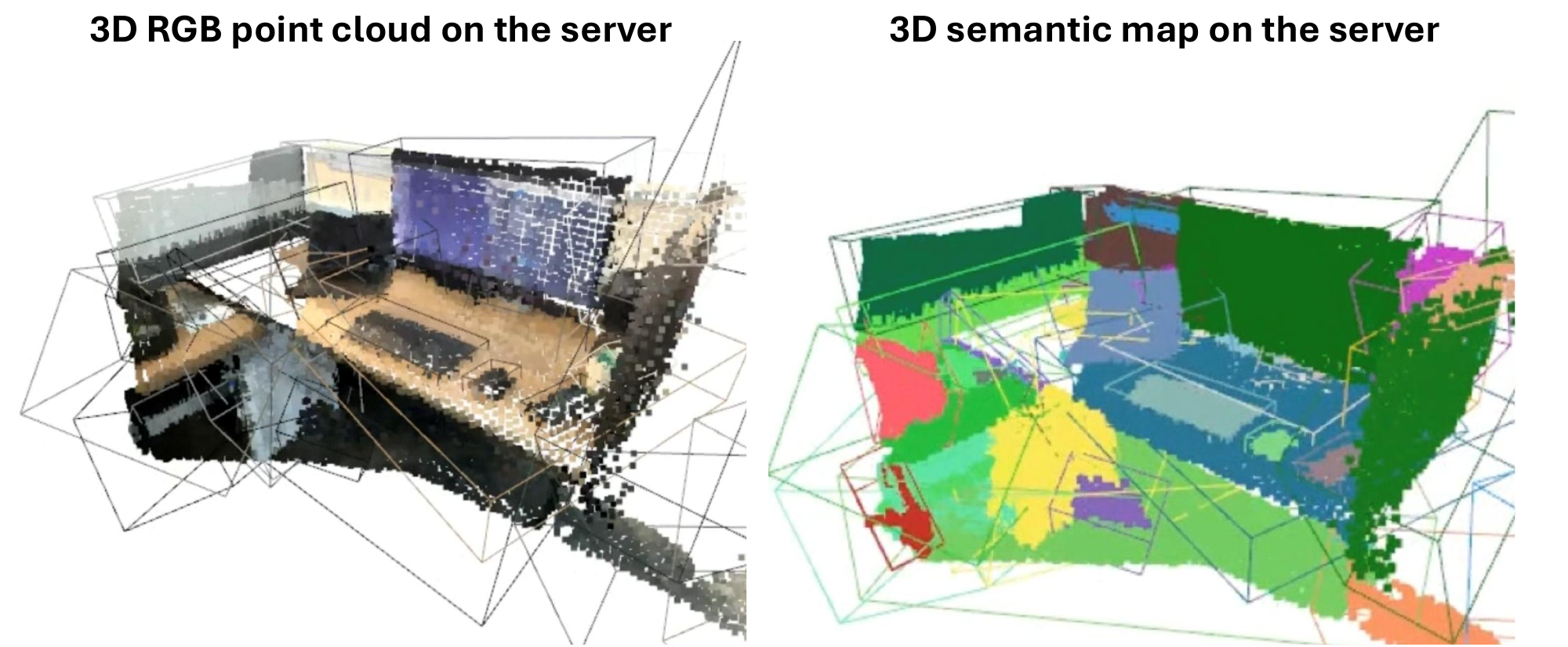}
        \caption{Semantic mapping: Left: Reconstructed color point cloud on the server. Right: 3D semantic map represented as object-based point cloud.}
        \label{fig:ipad_server}
    \end{subfigure}

    \vspace{0.1in}
    
    \begin{subfigure}[b]{0.9\columnwidth} 
        \centering
        \includegraphics[width=1.1\columnwidth]{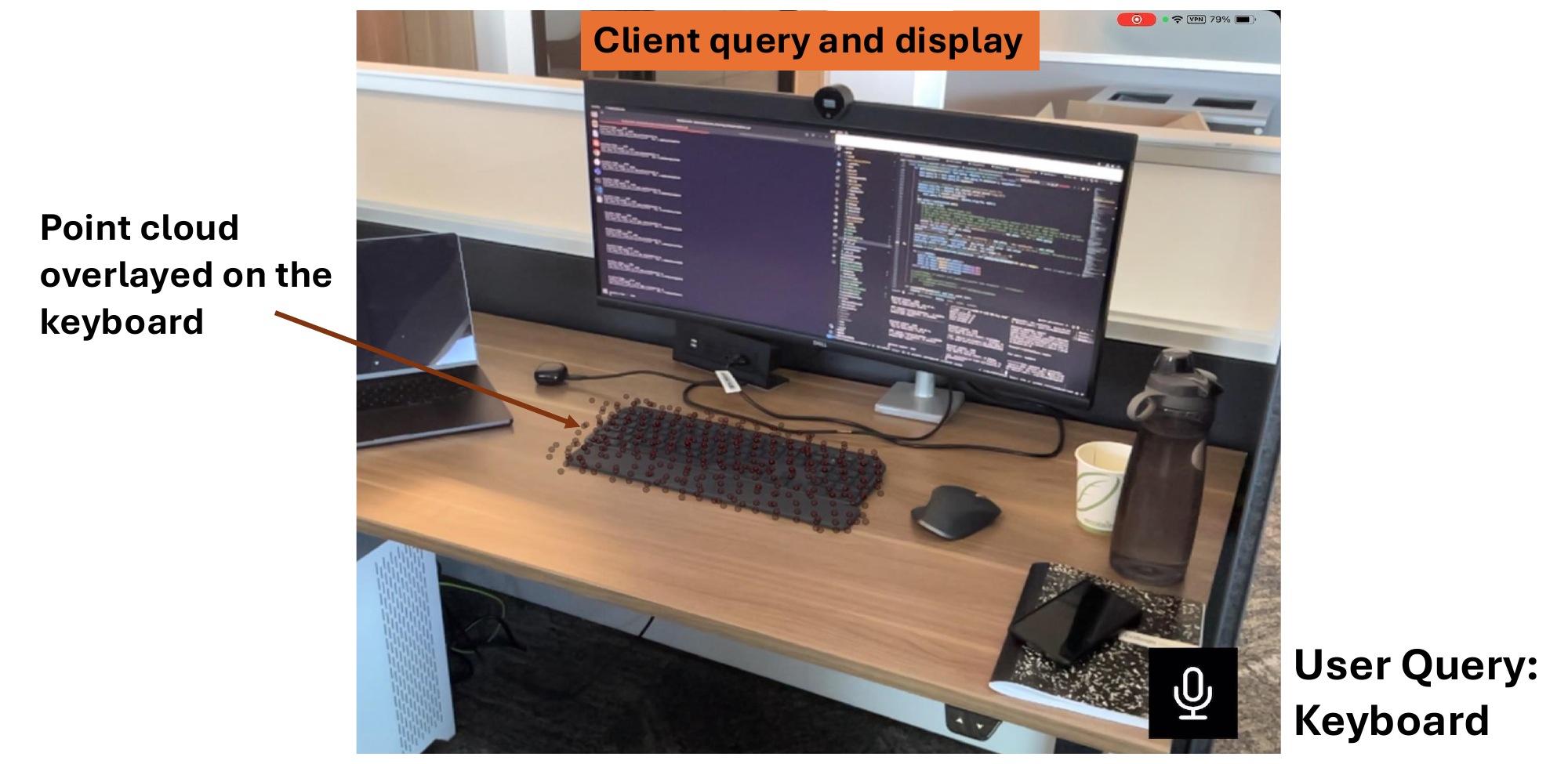} 
        \caption{iPad client view: The user presses the mic button to submit a voice query for a keyboard. The iPad receives the keyboard point cloud and overlays it on the physical keyboard.}
        \label{fig:ipad_client}
    \end{subfigure}
    \vspace{-0.1in}
    \caption{\name deployed end-to-end: (a) Overview of semantic 3D scene mapping. (b) Query output. This figure is created with an iPad as the client. The accompanying video shows a demonstration with a Quest 3 as the client.}
    \label{fig:ipad_whole}
    \vspace{-0.2in}
\end{figure}

\section{Deployment Case Study: Quest and iPad Prototypes}
\label{ipad}

To validate \name with real sensors, we built end-to-end prototypes on two consumer platforms: a Meta Quest~3~\cite{quest3} mixed-reality headset and an Apple iPad~Pro~\cite{ipad}. On the iPad, the client obtains RGB, depth, and pose through ARKit~\cite{arkit}; on the Quest~3, it obtains RGB and depth through Meta's Passthrough Camera and Depth APIs and pose from headset tracking, all in Unity~\cite{metapca, metadepth}. Deploying on an actual XR headset in addition to the iPad shows that \name's lightweight device-side runtime ports across mobile XR sensing stacks.

Each client streams a subset of captured frames to the server to match mapping throughput and exposes \name's tunable interface. The iPad transmits H.264 RGB ($720\times1280$, 5\,Mbps) with pose and depth ($144\times256$); the Quest~3 transmits native-resolution RGB with pose and depth ($320\times320$). For spatial queries, the client streams audio to the server, which transcribes it, queries the semantic map, and returns matching object-level point clouds that the client overlays in the user's view.

Although neither device exposes fine-grained power telemetry (hence our Jetson-based power characterization, \cref{eval:eval_sys}), these deployments demonstrate end-to-end operation on real mobile hardware, including an actual XR headset, with live sensors and interactive queries. A demonstration video of the Quest~3 deployment is included as supplementary material. \cref{fig:ipad_whole} shows the iPad deployment: the server incrementally builds the semantic map (\cref{fig:ipad_server}) and the client overlays queried objects in the user's view (\cref{fig:ipad_client}).

\section{Discussion}
\label{sec:discussion}

\subsection{Applicability Across Semantic Mapping Pipelines}
\label{disc:generalize}
\name's system innovations (\cref{tab:sys_req_innov}) are not tied to a specific perception model; they require only that the backend produce persistent, discrete object-level semantic state that can be incrementally maintained. Backends that already maintain per-object representations (e.g.,~\cite{conceptgraph, conceptfusion, clio, onlineAnySeg}) can adopt \name's abstractions directly by promoting objects to independently managed system state. Pipelines that aggregate object inference into monolithic scene representations~\cite{open-fusion,multimodal3Dfusion:ISMAR24} must first expose persistent object entities, requiring more substantial restructuring.



\subsection{\rev{Query Scope}}
\label{disc:query_scope}
\rev{\name is evaluated on object-grounding queries, which retrieve the scene objects that best match a natural-language description. The object-level organization extends to other query classes through targeted additions rather than a redesign: multi-object and relational queries via an external reasoning agent operating over the object map, affordance queries via additional per-object attributes, and free-space or global-layout queries via complementary representations such as occupancy.}


\subsection{Relationship to Model-Level Optimization}
Model-level optimization, such as quantization, pruning, and distillation, is orthogonal to \name's object-level system organization. Because the foundation models run on the server, such optimization reduces server-side mapping latency and does not reduce device power. Object-level system organization is complementary in the way
it reduces computation, communication, and memory cost, affecting latency and device power.

\subsection{Future Work}

Future directions include server-side scalability to multiple concurrent devices, support for dynamic environments with moving objects and multi-user collaborative mapping, and extending the object-level interface to richer geometric representations such as per-object meshes with non-uniform detail. \name's configurable knobs (\cref{sys:gen_tunable}) also motivate autotuning and dynamic control that build on existing autotuning frameworks~\cite{opentuner, approxtuner, slambooster}, automatically adapting resource-quality parameters to application and network conditions instead of manual configuration. 
Agentic integration is another direction: humans could issue complex spatial
tasks that an agent resolves by issuing spatial queries against SemanticXR's map and acting on the returned objects.
Privacy-aware design for device-cloud XR systems remains an important open problem. Quantitative power and thermal characterization on commercial XR headsets, together with evaluation in outdoor environments, would further validate deployment generality.
\section{Related Work}
\label{sec:relatedwork}

Prior work has studied semantic mapping primarily as an algorithmic perception problem, while cloud offloading in XR has largely focused on rendering or isolated perception tasks. In contrast, deploying semantic mapping as a low-power, long-running XR service requires managing communication, execution, and memory footprint across a device-cloud boundary, a problem that existing approaches do not address.

\subsection{Semantic Mapping in Robotics and XR}


A large body of work has explored open-vocabulary semantic mapping by embedding image- or text-derived semantics into 3D representations. Representative systems include Clio~\cite{clio}, SpatialLM~\cite{spatiallm}, OpenFusion~\cite{open-fusion}, ConceptFusion~\cite{conceptfusion}, OpenMask3D~\cite{openmask3d}, ConceptGraph~\cite{conceptgraph}, and One Map to Find Them All~\cite{onemaptofindthemall}. These approaches differ in perception models, representations, and update strategies, and collectively demonstrate the feasibility of constructing semantic maps from RGB-D observations. Many support online RGB-D processing~\cite{clio,open-fusion}, while others operate offline over preprocessed point clouds. However, these systems target algorithmic quality under server-class hardware assumptions, whether through task-driven scene representations~\cite{clio} or monolithic scene-level representations~\cite{open-fusion, multimodal3Dfusion:ISMAR24}, rather than system organization for device-cloud deployment under XR constraints. 

\rev{Unlike these open-vocabulary systems, a related line of robotics work builds real-time hierarchical 3D scene graphs from closed-vocabulary perception, including Hydra~\cite{hydra, hydra_foundations} and SceneGraphFusion~\cite{sceneGraphFusion}. Because this perception is lightweight, these systems run on a single machine and do not address the device-cloud split that foundation-model perception forces on mobile XR. 
Clio~\cite{clio} extends this scene-graph line toward open-vocabulary, task-driven mapping and is the closest comparison among these systems. As shown in \cref{tab:quality_comparison}, \name matches Clio-batch quality without offline post-processing and improves on Clio-online in the latency and quality trade-off.}
Separately, 3D instance segmentation methods~\cite{embodSAM,onlineAnySeg,ovir3d} focus on geometric shape decomposition rather than persistent, queryable semantic state and do not target device-cloud deployment.
In contrast, \name treats existing semantic mapping pipelines as backends and elevates objects to first-class units of device-cloud system design, governing communication, execution, and memory footprint across the device-cloud boundary (\cref{tab:sys_req_innov}).

\subsection{VLM-based Assistants and Scene Understanding}
Recent XR systems have explored natural language interaction using VLM-based assistants such as Google's Project Astra~\cite{astra} and XaiR~\cite{xair}. These systems reason over visual observations together with short-term textual or embedding-based memory~\cite{astra}, or attach per-frame features to pre-built geometry without distinguishing individual objects~\cite{xair}. Neither builds a persistent, queryable 3D semantic map with explicit object identities. \name addresses an orthogonal problem: how to build and manage such a map across a device-cloud boundary under the power, bandwidth, and memory constraints of mobile XR.

\subsection{Cloud Offloading in XR}
Several prior works have explored cloud offloading for XR to address the compute and power limitations of mobile devices, including offloading components such as head pose estimation, visual tracking, and rendering \cite{remotevio, cloudxr, xrgo, renderfusion, furion, slimslam, slamshare, adaPang}. These systems primarily focus on latency-sensitive rendering pipelines or geometric reconstruction, and do not consider the challenges associated with maintaining and querying a persistent semantic map. In addition, offloading AI workloads such as object detection and segmentation has been widely studied~\cite{accumo, arise, elf, edgeAssistedObjectDetection, marvel, vips, rao}. While some of these systems operate at object granularity, they typically target task-specific domains (e.g., object detection over a limited and predefined set of categories) and do not address the requirements of open-vocabulary semantic mapping. Techniques such as keyframe sampling reduce bandwidth or server compute, but do not address the system-level challenges introduced by offloading semantic mapping in XR, including maintaining persistent semantic state and robustness to network variability.

\subsection{Semantic Maps on Neural Representations}
Recent work has explored semantic scene representations based on neural fields, such as NeRFs~\cite{nerf} augmented with language grounding~\cite{lerf}, and Gaussian-based representations~\cite{gaussian} extended for semantic queries~\cite{langsplat,sgsslam}. While effective for dense reconstruction and semantic querying, these representations typically require optimization over accumulated observations and are not yet designed for incremental, low-latency updates under the power, bandwidth, and memory constraints of mobile XR systems.

\textit{Taken together,} prior work provides strong semantic mapping backends and cloud-XR mechanisms, but does not address how to manage communication, execution, and memory footprint for semantic mapping across a device-cloud boundary under the power, bandwidth, and memory constraints of mobile XR. \name addresses this gap through object-level system organization that is compatible with a broad class of existing semantic mapping pipelines.




\section{Conclusion}
\label{sec:conclusion}
We presented \name, the first device-cloud system for real-time, open-vocabulary semantic mapping and querying within the power, bandwidth, and memory constraints of mobile XR. Our key insight is that semantically identifiable objects can serve as first-class units of device-cloud system design, governing how the system communicates, executes, and manages memory across the device and the server. This object-level system organization enables a family of innovations (\cref{tab:sys_req_innov}) that collectively address the requirements of XR device-cloud semantic mapping: object-level parallelism and geometry downsampling improve server-side mapping latency by 2.2$\times$ at equivalent semantic quality, object-level depth-mapping co-design reduces upstream bandwidth by $\sim$90\% with a small quality loss, and an object-level sparse local map with incremental updates and update prioritization enables sub-100 ms query latency for up to 10{,}000 objects and holds 50{,}000 objects within 500 MB, with downstream bandwidth proportional to map changes rather than total scene size. The system adds only $\sim$2\% device power over the idle power. Because \name operates on object-level semantic state rather than model-specific internals, its system innovations generalize across any semantic mapping pipeline that produces persistent per-object representations.

\acknowledgments{
We thank the anonymous reviewers for their constructive feedback. This work was supported in part by the National Science Foundation (NSF) under grants 2120464, 2217144, and 2454014, and by an internship at NVIDIA.
}

\bibliographystyle{abbrv-doi}
\bibliography{new_ref}
\end{document}